\documentclass[aps,prb,twocolumn,superscriptaddress,showpacs]{revtex4}
\usepackage{graphicx}
\usepackage{latexsym}
\usepackage{amssymb}
\usepackage{amsmath}
\usepackage{amsfonts}
\usepackage{bm}
\usepackage{multirow}
\usepackage{color}
\usepackage{comment}

\newcommand{\beq}{\begin{equation}}
\newcommand{\eeq}{\end{equation}}
\newcommand{\beqn}{\begin{eqnarray}}
\newcommand{\eeqn}{\end{eqnarray}}

\DeclareMathAlphabet{\mathbbold}{U}{bbold}{m}{n}

\makeatletter
\newcommand\xleftrightarrow[2][]{%
\ext@arrow 9999{\longleftrightarrowfill@}{#1}{#2}}
\newcommand\longleftrightarrowfill@{%
\arrowfill@\leftarrow\relbar\rightarrow} \makeatother

\begin{document}

\title{A candidate Theory for the ``Strange Metal" phase at Finite Energy Window}

\author{Xiaochuan Wu}
\affiliation{Department of Physics, University of California,
Santa Barbara, CA 93106, USA}

\author{Xiao Chen}
\affiliation{Kavli Institute of Theoretical Physics, Santa
Barbara, CA 93106, USA}

\author{Chao-Ming Jian}
\affiliation{ Station Q, Microsoft Research, Santa Barbara,
California 93106-6105, USA} \affiliation{Kavli Institute of
Theoretical Physics, Santa Barbara, CA 93106, USA}

\author{Yi-Zhuang You}
\affiliation{Department of Physics, Harvard University, Cambridge,
MA 02138, USA}

\author{Cenke Xu}
\affiliation{Department of Physics, University of California,
Santa Barbara, CA 93106, USA}

\date{\today}
\begin{abstract}

We propose a lattice model for strongly interacting electrons with
the potential to explain the main phenomenology of the strange
metal phase in the cuprate high temperature superconductors. Our
model is motivated by the recently developed ``tetrahedron" rank-3
tensor model that mimics much of the physics of the better-known
Sachdev-Ye-Kitaev (SYK) model. Our electron model has the
following advantageous properties: {\it 1.} it only needs one
orbital per site on the square lattice; {\it 2.} it does not
require any quenched random interaction; {\it 3.} it has local
interactions and respects all the symmetries of the system; {\it
4.} the soluble limit of this model has a longitudinal DC
resistivity that scales linearly with temperature within a finite
temperature window;
{\it 5.} again the soluble limit of this model has a fermion
pairing instability in the infrared, which can lead to either
superconductivity or a ``pseudogap" phase.
The linear$-T$ longitudinal resistivity and the pairing
instability originate from the generic scaling feature of the SYK
model and the tetrahedron tensor model.

\end{abstract}

\maketitle

\section{Introduction}

Non-fermi liquid (NFL) state represents a family of exotic
metallic states that do not have long-lived quasi-particles, and
hence behave fundamentally differently from the standard Landau
Fermi liquid
theory~\cite{nfl1,polchinskinfl,nayaknfl1,nayaknfl2,nematicnfl,nfl2,nfl3,nfl4,nfl5,nfl6,nflqmc}.
The NFLs usually occur at certain quantum critical point in
itinerant fermion systems, and the quantum critical fluctuations
couple strongly with the fermions and hence ``kill" the
quasiparticles. But the most well-known (yet poorly understood)
NFL, the ``strange metal" phase at the optimal doping of the
cuprate high temperature superconductors, seems more generic than
the byproduct of a certain quantum critical point, because its
anomalous temperature dependence of longitudinal DC resistivity
($\rho \sim T$) persists up to a rather high temperature in the
phase diagram~\cite{linear1,linear2,linear3,linear4,Varma1989},
which is presumably much higher than the ultraviolet cut-off of
any possible quantum critical point in the system. However, like
many other
NFLs~\cite{nflpairing1,nflpairing2,nflpairing3,nflpairing4,nflpairing5,nflpairing6,inter},
the strange metal phase is also preempted by a dome of ``ordered
phase" with pair condensate of fermions (high $T_c$
superconductivity) at low temperature. Thus the strange metal
phase is more fundamental than the superconductor phase itself: it
is the ``parent state" of the high $T_c$ superconductor, just like
the Fermi liquid is the parent state (or normal state) of
conventional BCS superconductors. And we had better view this
parent state as a generic non-Fermi liquid state, instead of a
quantum critical behavior.

A series of toy models for NFL, despite their relatively unnatural
forms, seem to capture the key universal features mentioned above.
These models are the so-called Sachdev-Ye-Kitaev (SYK) model and
its
generalizations~\cite{SachdevYe1993,Kitaev2015,Sachdev2015,Polchinski2016,MaldacenaStanford2016,Witten2016,Klebanov2016,Gross2017}.
{\it 1.} the fermion Green's function in these models has a
completely different scaling behavior from the noninteracting
fermions in the infrared limit, thus it has no quasi-particle and
by definition is a NFL. {\it 2.} it was found that the SYK model
has marginally relevant ``pairing instability" just like the
ordinary Fermi liquid state~\cite{Xu2017,pairex}, which is again
consistent with one of the universal features of the NFLs observed
experimentally. {\it 3.} Recently measured charge density
fluctuation of the strange metal~\cite{abbamonte} agrees with the
unique scaling behavior of the SYK model~\cite{SachdevYe1993}.
{\it 4.} Last but not least, recently a generalization based on
the SYK model has shown linear-$T$ resistivity for a large
temperature window, and the scaling behavior of the SYK model is
the key for the linear-$T$ resistivity~\cite{Song2017} (similar
effect can be achieved in models with large$-N$ generalization of
the electron-phonon coupling~\cite{phonon1,phonon2,phonon3}). All
these developments suggest that some version of the SYK model and
its generalizations may indeed have to do with the strange metal
phase.

More often than not, an exactly soluble model has to sacrifice
reality to some extent by making some artificial assumptions. To
ensure its solubility, the original SYK model has the following
necessary ingredients that make it unlikely to be directly related
to the cuprates: {\it 1.} It needs an all-to-all four-fermion
interaction, while a natural Hamiltonian for a real condensed
matter system usually has local interactions only; {\it 2.} The
four-fermion interaction is fully random with a Gaussian
distribution, which is also far from the real system. {\it 3.} So
far the NFL models constructed based on generalizations of the SYK
model all have a large number of fermion states on each unit-cell
of the lattice with a fully random all-to-all intra unit-cell
interaction~\cite{Gu2017,Gu2017b,yao2017,xu2017b,Song2017,patel2017,berg2018},
while the common wisdom is that the cuprate materials only have
one active $d-$orbital on each copper site.

\begin{figure}[tbp]
\begin{center}
\includegraphics[width=200pt]{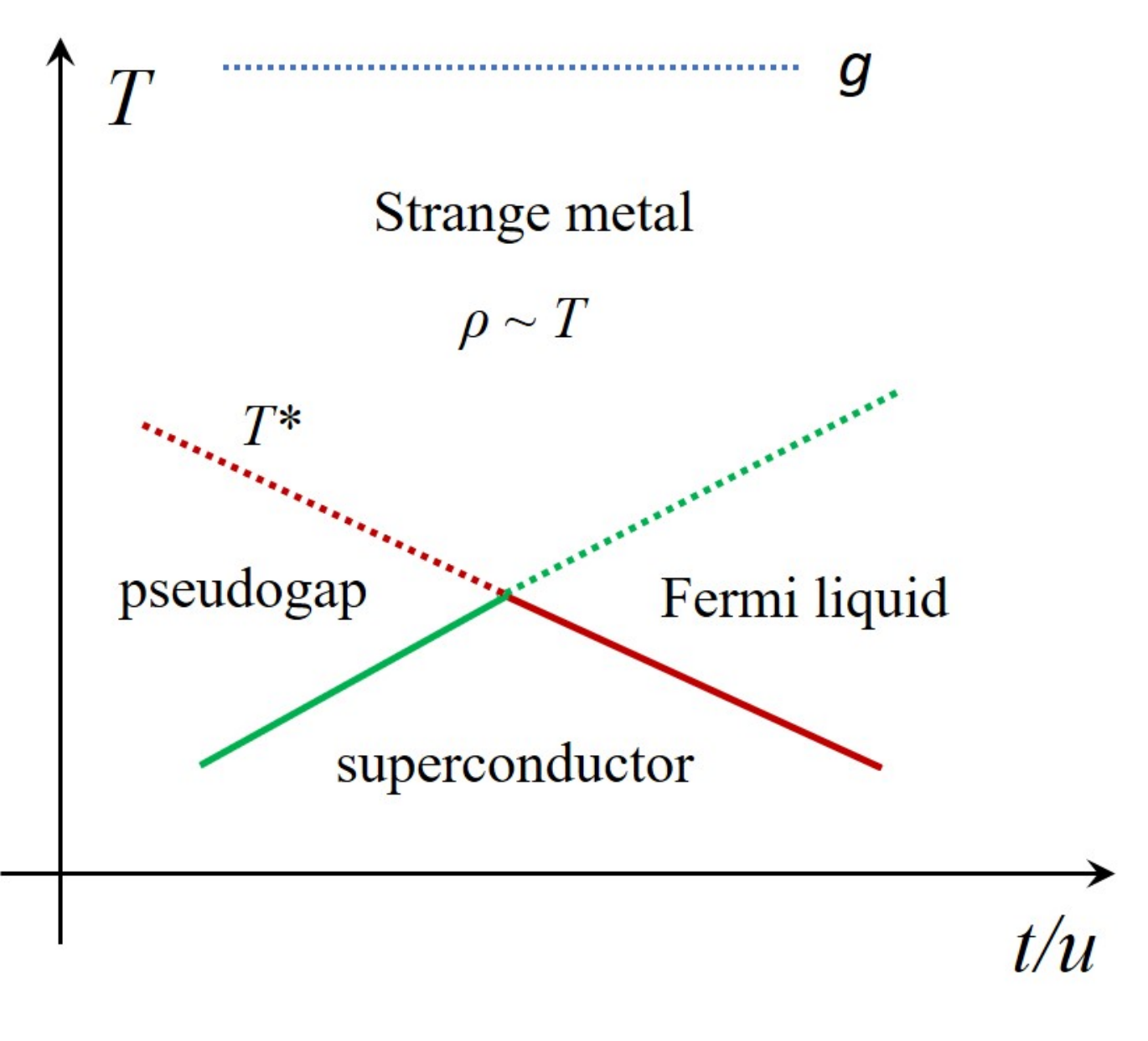}
\caption{The schematic phase diagram of our Hamiltonian
Eq.~\ref{Hs} or Eq.~\ref{Hs2} plus single particle hopping
parametrized by $t$ and nearest neighbor perturbation $H_u$
(Eq.~\ref{H1}) with coefficient $u$. The strange metal phase is
dominated by Eq.~\ref{Hs} or Eq.~\ref{Hs2} only, and is
characterized by the non fermi liquid behavior and an anomalous
linear$-T$ scaling of the DC resistivity. The pseudogap crossover
temperature scale $T^\ast$ is given by Eq.~\ref{E}. The exact
phase boundaries need further calculations.} \label{pd}
\end{center}
\end{figure}

In this work we will construct two lattice models for strongly
interacting electrons that are still motivated by the SYK physics,
but are much closer to real systems. {\it 1.} Our models only need
one orbital per unit-cell on the square lattice; {\it 2.} Our
models have no quenched randomness; {\it 3.} Our models still
capture the most desired physics of the SYK model, such as the
linear$-T$ scaling of the longitudinal DC resistivity, and pairing
instability in the infrared. In the soluble limit, the solution of
our model is identical to the SYK model, thus our analytical
results largely rely on the known solution of the SYK model in for
instance Ref.~\onlinecite{Sachdev2015}. But we will also check our
analytical predictions based on the soluble limit by exact
diagonalization of the minimal and most realistic version of our
model away from the soluble limit, on a finite system. The phase
diagram of our proposed model for the physics near the strange
metal phase including the low energy phases induced by different
perturbations considered in this paper are plotted in
Fig.~\ref{pd}.

It was shown previously for the Sachdev-Ye model, that away from
the exactly soluble large$-N$ limit~\footnote{Actually the
original Sachdev-Ye model requires two parameters, $N$ and $M$, be
taken infinite. Here for simplicity we use large$-N$ to represent
both limits.}, the SYK scaling still persists at finite energy
scale (for example finite temperature), while instabilities due to
$1/N$ corrections emerge at low energy which are suppressed
(sub)exponentially with increasing$-N$~\cite{sachdev2001}.
Although the exactly soluble version of our models still requires
some large$-N$ limit, by evaluating the next order diagrams, we
argue that for finite$-N$, the scaling behavior of the large$-N$
limit may still apply to an intermediate energy or temperature
window, which is where the strange metal phase was observed in
real systems.

\section{The Hamiltonian}

Let us first write down the most important term of the interacting
electron Hamiltonian that we will study on the square lattice:
\beqn H &=& \sum_j H_j, \cr\cr H_j &=& U \hat{n}_{j}^2 +
\sum_{\hat{e} = \hat{x}, \hat{y}}J \left( \vec{S}_{j} \cdot
\vec{S}_{j + \hat{e}} - \frac{1}{4} \hat{n}_{j}\hat{n}_{j +
\hat{e}} \right) \cr\cr &-& K \left( \epsilon_{\alpha\beta}
\epsilon_{\gamma\sigma} c^\dagger_{j, \alpha} c^\dagger_{j +
\hat{x} + \hat{y}, \beta} c_{j+\hat{y}, \gamma} c_{j + \hat{x},
\sigma} + H.c. \right), \label{Hs} \eeqn where
$\epsilon_{\alpha\beta}$ is an $2\times 2$ antisymmetric matrix in
the spin space. Other terms, such as single particle hopping, will
later be treated as perturbations. We will study this model with a
fixed particle density both analytically and numerically.
$\hat{n}_{j} = \hat{n}_{j, \uparrow} + \hat{n}_{j, \downarrow}$ is
the total electron number on site $j$, $\vec{S}_{j} = \frac{1}{2}
c^\dagger_{j} \vec{\sigma} c_{j}$ is the spin operator. Besides
the standard charge density and spin interactions, we also turned
on a ``ring exchange" term with coefficient $K$, which takes a
spin singlet pair of electrons on two diagonal sites of a
plaquette to the two opposite diagonal sites of the same
plaquette. This Hamiltonian preserves the square lattice symmetry
(because this interaction only has parity-even and spin singlet
pairing between fermions), and also spin SU(2) symmetry.

We will try to make connection between Eq.~\ref{Hs} and the SYK
physics. As we explained previously, many necessary ingredients of
the original SYK model are not very realistic. Instead of directly
using the SYK model, our construction Eq.~\ref{Hs} is motivated by
the randomness-free ``tetrahedron" model (or the so-called rank-3
tensor model)~\cite{Gurau,Witten2016,Klebanov2016}: \beqn H^t_1 =
\frac{g}{(N_a N_b N_c)^{1/2}} c^\dagger_{a_1 b_1 c_1}
c^\dagger_{a_2 b_2 c_1} c_{a_1 b_2 c_2} c_{a_2 b_1 c_2}.
\label{Ht} \eeqn $a_1, a_2$ = $1 \cdots N_a$, $b_1, b_2 = 1 \cdots
N_b$, and $c_1, c_2 = 1 \cdots N_c$. This model has a
$\mathrm{U}(N_a) \times \mathrm{U}(N_b) \times \mathrm{O}(N_c)$
symmetry.
It was shown in the literature that, in the large $N_i$ limit, the
dominant contribution to the Fermion Green's function comes from a
series of ``melon Feynman diagrams", which can be summed
analytically by solving the Schwinger-Dyson equation.

To make connection to electron systems, the first step is to
modify the tetrahedron model as follows: \beqn H^t_{2} &=& -
\frac{g}{(N_a N_b N_c)^{1/2}} \cr\cr &\times& \mathcal{J}_{c_1,
c_1'} \mathcal{J}_{c_2, c_2'} c^\dagger_{a_1 b_1 c_1}
c^\dagger_{a_2 b_2 c_1'} c_{a_1 b_2 c_2} c_{a_2 b_1 c_2'},
\label{Ht2}\eeqn where $\mathcal{J}$ is the antisymmetric matrix
associated with the Sp($N_c$) group, and $\mathcal{J}_{ab} c_a c_b
$ forms a Sp($N_c$) singlet. The total symmetry of this model is
now $\mathrm{U}(N_a) \times \mathrm{U}(N_b) \times
\mathrm{Sp}(N_c)$. The solubility of this model is unchanged from
Eq.~\ref{Ht} in the large$-N_i$ limit, and the single particle
Green's function in this limit is identical to the
disorder-averaged Green's function of the SYK
model~\cite{Sachdev2015}: \beqn
G\left(\tau\right)&=&-\mathcal{B}\left(\theta\right)
e^{-2\pi T\mathcal{E}\tau}\sqrt{\frac{\pi T}{2g\sin\left(\pi T\tau\right)}}, \label{green3} \\
G\left(i\omega\right)_{T=0}&=&\frac{\mathcal{B}\left(\theta\right)}{\sin\left(\frac{\pi}{4}+\theta\right)}
\frac{e^{-i\textrm{sgn}\left[\omega\right]
\left(\frac{\pi}{2}+\theta\right)}}{\left|2g\omega\right|^{\frac{1}{2}}},
\label{green} \eeqn where a real angle parameter
$-\frac{\pi}{4}<\theta<\frac{\pi}{4}$ and the spectral asymmetry
$\mathcal{E}$ have been introduced. Both parameters depend on the
charge density, and they are related to each other by \beqn
e^{2\pi\mathcal{E}}=\frac{\sin\left(\frac{\pi}{4}+\theta\right)}{\sin\left(\frac{\pi}{4}-\theta\right)}.
\eeqn The angle $\theta=0$ corresponds to the case of
half-filling. By solving consistent equations with the same method
as Ref.~\onlinecite{Sachdev2015}, the coefficient $\mathcal{B}$ is
found to be \beqn \mathcal{B}\left(\theta\right)=
\left(\frac{1}{\pi\cos\left(2\theta\right)}\right)^{\frac{1}{4}}\sin\left(\frac{\pi}{4}+\theta\right).
\eeqn In Eq.~\ref{green3}, we have assumed $0< \tau <\beta$ in the
Green's function, and the Green's function with $- \beta < \tau <
0$ is determined by the standard relation
$G\left(\tau+\beta\right)=-G\left(\tau\right)$.

Now we can draw connection between the modified tetrahedron model
Eq.~\ref{Ht2} and our original model Eq.~\ref{Hs}. When $U = K =
\eta J/2$ ($\eta = \pm 1$), the total Hamiltonian Eq.~\ref{Hs} is
equivalent to the following model with $N = 3$ and $M = 2$: \beqn
&& H = \sum_{j} \sum_{r, r' = - (N-1)/2 }^{(N-1)/2} \
\sum_{\alpha,\beta, \gamma,\sigma = 1}^{M} - \frac{g
\eta_{r,r'}}{N \sqrt{M}}  \cr\cr &\times& \mathcal{J}_{\alpha
\beta} \mathcal{J}_{\gamma\sigma} c^\dagger_{j_x, j_y, \alpha}
c^\dagger_{j_x + r, j_y + r', \beta} c_{j_x, j_y + r',\gamma}
c_{j_x + r, j_y, \sigma}. \label{H}\eeqn Just like the tetrahedron
model Eq.~\ref{Ht2}, every fermion still carries three indices:
the Sp($M$) spin, the $x-$coordinate, and $y-$coordinate. We will
consider and numerically study two versions of the models with
$\eta_{r,r'} = + 1$ uniformly (when $N = 3$, $M = 2$ it
corresponds to $U = K = - J/2$) and $\eta_{r,r'} = (-1)^{r + r'}$
(which corresponds to $U = K = + J/2$) respectively. Here we allow
$J$ to take both signs. Although an antiferromagnetic order is
well-known in cuprates in the underdoped regime, ferromagnetism
has also been discussed in the overdoped regime~\cite{ferro}.

The minimal version of the model Eq.~\ref{H} with $N = 3$, $M=2$,
is identical to Eq.~\ref{Hs}, which should be analogous to the
case with $N_a = N_b = 3$ in Eq.~\ref{Ht2}. In analytical
calculations, we always take the thermodynamics limit first (the
sum of $j$ is taken on a square lattice with infinite size). Then
in the large$-N$ and large$-M$ limit, for both choices of
$\eta_{r,r'}$, the fermion Green's function is still dominated by
the ``melon diagrams", and hence the Schwinger-Dyson equations, as
well as their solutions, remain the same as models Eq.~\ref{Ht},
and Eq.~\ref{Ht2}: \beqn
G_{j,j^{\prime},\alpha,\beta}\left(\tau\right)=
G\left(\tau\right)\delta_{j,j^{\prime}}\delta_{\alpha,\beta},
\label{green2} \eeqn from which we can extract the fermion
spectral function (local density of states) \beqn
\rho_{f}\left(\omega\right)=\sqrt{\frac{1}{gT}}
\frac{\mathcal{B}\left(\theta\right)}{\sin\left(\frac{\pi}{4}+\theta\right)}
\textrm{Im}\left[\frac{ie^{-i\theta}}{2\pi}
\frac{\Gamma\left(\frac{1}{4}+\frac{\beta\left(\omega-\omega_{\mathcal{S}}\right)}{2\pi
i}\right)}{\Gamma\left(\frac{3}{4}+\frac{\beta\left(\omega-\omega_{\mathcal{S}}\right)}{2\pi
i}\right)}\right]. \label{spectral} \eeqn Here
$\omega_{\mathcal{S}}=2\pi\mathcal{E}T$. The Fermion Green's
function has a form of local quantum criticality, and the scaling
dimension of the fermion operator is $\Delta[c] = 1/4$.

We have introduced a fixed fermion density defined as \beqn
\mathcal{Q}=\frac{1}{M}\sum_{\alpha=1}^{M}\left\langle
c_{j,\alpha}^{\dagger}c_{j,\alpha}\right\rangle. \eeqn The value
of $\mathcal{Q}$ can be varied within the range $0<\mathcal{Q}<1$.
Using the same method as Ref.~\onlinecite{Sachdev2015}, the
relation between fermion density $\mathcal{Q}$ and the angle
parameter $\theta$ in the Green's function is found to be \beqn
\mathcal{Q}=\frac{1}{2}-\frac{\theta}{\pi}
-\frac{\sin\left(2\theta\right)}{4},\qquad-\frac{\pi}{4}<\theta<\frac{\pi}{4}.
\label{number and theta} \eeqn

The fact that the Fermion Green's function Eq.~\ref{green2}
remains localized in space is due to the fact that the Hamiltonian
Eq.~\ref{Hs} and Eq.~\ref{H} preserve the center-of-mass of the
electrons on the square lattice. Any nonzero fermion correlation
with a finite spatial separation would violate the center of mass
conservation, thus the Fermion Green's function is fully localized
in space. Single particle hopping will later be introduced as
perturbation, which breaks center-of-mass conservation and leads
to spatial correlation between fermions, and also charge
transport.

For finite $N$ and $M$, we need to estimate the corrections coming
from the subdominant Feynman diagrams. For any diagram, if we
evaluate it with the solution in the large$-N, M$ limit, it will
roughly lead to a ``marginal" correction, namely it will correct
the large$-N, M$ solution with a logarithmic function of infrared
cut-off, say the temperature. This is because in the large$-N, M$
soluble limit the coupling constant $g$ becomes marginal, since
the scaling dimension of the fermion operator is $1/4$.
Subdominant Feynman diagrams of SYK like models have been
carefully calculated in Ref.~\onlinecite{nextorder}, and the
result is consistent with our expectation. Thus we expect that any
subdominant diagram will {\it at most} lead to corrections with
the form $\sim 1/N^A 1/M^B (\log(\Lambda/T))^C$, where $A, B$ and
$C$ are all positive numbers. This diagram will hence become
significant only when \beqn T \leq \Lambda \exp(- c
N^{\frac{A}{C}}M^{\frac{B}{C}}), \eeqn where $\Lambda$ is the
ultraviolet cut-off of the system, which can be identified as $g$
in our model. Thus we expect the correction to the NFL solution is
suppressed rapidly with increasing $N$ and $M$, hence it is
possible that there is a finite energy window where the solution
Eq.~\ref{green2} applies. This is consistent with the expectation
for the original Sachdev-Ye model away from the exactly soluble
limit~\cite{sachdev2001}. Away from the exactly soluble limit, the
ground state has no finite entropy density.

\section{Properties of the NFL}

\subsection{Longitudinal Conductivity}

Assuming Eq.~\ref{green2} applies to a finite energy window, we
can use it to compute quantities at finite temperature within such
energy window. Because Eq.~\ref{Hs} conserves the center of mass
of the electrons, it is incapable of transporting electric charge.
More formally, this interaction term does not couple to the zero
momentum component of the external electromagnetic field,
analogous to models studied previously with center of mass
conservation~\cite{ring1,ring2}. Thus the single particle hopping
term is still responsible for charge transport. In cuprates both
the nearest neighbor and second neighbor hoppings are
important~\cite{binding}. In the soluble large$-N,M$ limit, we
formally generalize the electric current density to the following
form: \beqn J_{x} &=& \frac{1}{\sqrt{NM}} \left( \sum_{\alpha}
itc_{j,\alpha}^{\dagger}c_{j+ \hat{x},\alpha} +
\sqrt{\frac{N-1}{2}} itc_{j,\alpha}^{\dagger}c_{j + \hat{x} \pm
\hat{y},\alpha} \right) \cr\cr &+& H.c. \label{current}\eeqn This
electric current density can be derived by designing a
corresponding single electron hopping term in the large$-N,M$
limit (which involves both nearest and second neighbor hopping),
and couple it to the external electromagnetic field.

Assuming the solution in the large$-N,M$ limit Eq.~\ref{green2}
applies to a finite energy window of the system, then according to
the Kubo formula, the central task is to calculate the retarded
current-current correlation function. The imaginary-time
correlation function is defined as
$C\left(J,J;\tau\right)=\left\langle
\mathbb{T}_{\tau}J\left(\tau\right)J\left(0\right)\right\rangle$.
We find $ \langle J_{x} \ J_{y} \rangle$ correlation vanishes due
to the symmetry of the model, and the leading order nonzero
contribution to $\langle J_{x} \ J_{x} \rangle$ takes the form
$C\left(J,J;\tau\right)=- 2t^{2}
G\left(\tau\right)G\left(-\tau\right)$. Then we Fourier transform
$C\left(J,J;\tau\right)$ to obtain the correlation function in the
Matsubara frequency space: \beqn C\left(J,J;i\omega_{n}\right)=
2t^{2}\int_{\delta}^{\beta-\delta}d\tau
e^{i\omega_{n}\tau}G\left(\tau\right)G\left(\beta-\tau\right),
\eeqn where we have regulated the integral by introducing a small
positive cut-off $\delta$. After removing the divergent term
$\log\delta$ (which does not contribute to the real part of the
conductivity), we obtain the analytically continued correlation
function \beqn
C\left(J,J;z\right)=-2\frac{t^{2}}{g}\mathcal{B}{}^{2}
e^{-2\pi\mathcal{E}}\psi\left(\frac{1}{2}+\frac{\beta z}{2\pi
i}\right), \eeqn where
$\psi\left(z\right)=\frac{d}{dz}\log\Gamma\left(z\right)$ is the
polygamma function, and the complex frequency $z$ satisfies
$\textrm{Im}z>0$. The function $C\left(J,J;i\omega_{n}\right)$ can
be obtained by setting $z\rightarrow i\omega_{n}$ on the above
expression, and the retarded/advanced correlation function
$C^{R/A}\left(J,J;\omega\right)$ is obtained by taking
$z\rightarrow\omega\pm i0^{+}$. Finally, using the relation
$\sigma\left(\omega\right)=\frac{1}{i\omega}C^{R}\left(J,J;\omega\right)$,
we find the real part of the optical conductivity \beqn
\textrm{Re}\sigma\left(\omega\right)= \frac{\sqrt{\pi}
t^{2}}{4gT}\varUpsilon_{\sigma}\left(\mathcal{Q},\omega/T\right),
\label{theocond} \eeqn where \beqn
\varUpsilon_{\sigma}\left(\mathcal{Q},\omega/T\right) =
\sqrt{\cos\left(2\theta\left(\mathcal{Q}\right)\right)}
\frac{\tanh\left(\omega/2T\right)}{\omega/2T} \eeqn is the scaling
function of conductivity. From another perspective,
$\varUpsilon_{\sigma}$ can also be computed from the convolution
of the scaling function of the fermion spectral function
$\rho_{f}$ in Eq.~\ref{spectral}.

By our definition, $\varUpsilon_{\sigma}$ depends on both the
fermion density $\mathcal{Q}$ and the ratio $\omega/T$. The
$\mathcal{Q}$-dependence of the conductivity is contained in the
coefficient $\sqrt{\cos\left(2\theta\right)}$ in the scaling
function $\varUpsilon_{\sigma}\left(\mathcal{Q},\omega/T\right)$,
and the function $\theta\left(\mathcal{Q}\right)$ can be obtained
by inverting Eq.~\ref{number and theta}. The half-filling
$\theta=0$ gives the maximum conductivity, as one would naively
expect. Once we fix the ratio $\omega/T$ (for example the DC limit
with $\omega/T = 0$), the longitudinal conductivity
$\sigma(\omega, T)$ is proportional to $1/T$, which is the most
important phenomenon of the strange metal phase.

In the calculation above we have assumed that the correlation
function between current operators factorizes into a product of
two Fermion Green's functions. This is true in the large$-N,M$
limit using the current operator Eq.~\ref{current}, and the
expression Eq.~\ref{theocond} is exact in this limit.

\begin{figure}[tbp]
\begin{center}
\includegraphics[width=255pt]{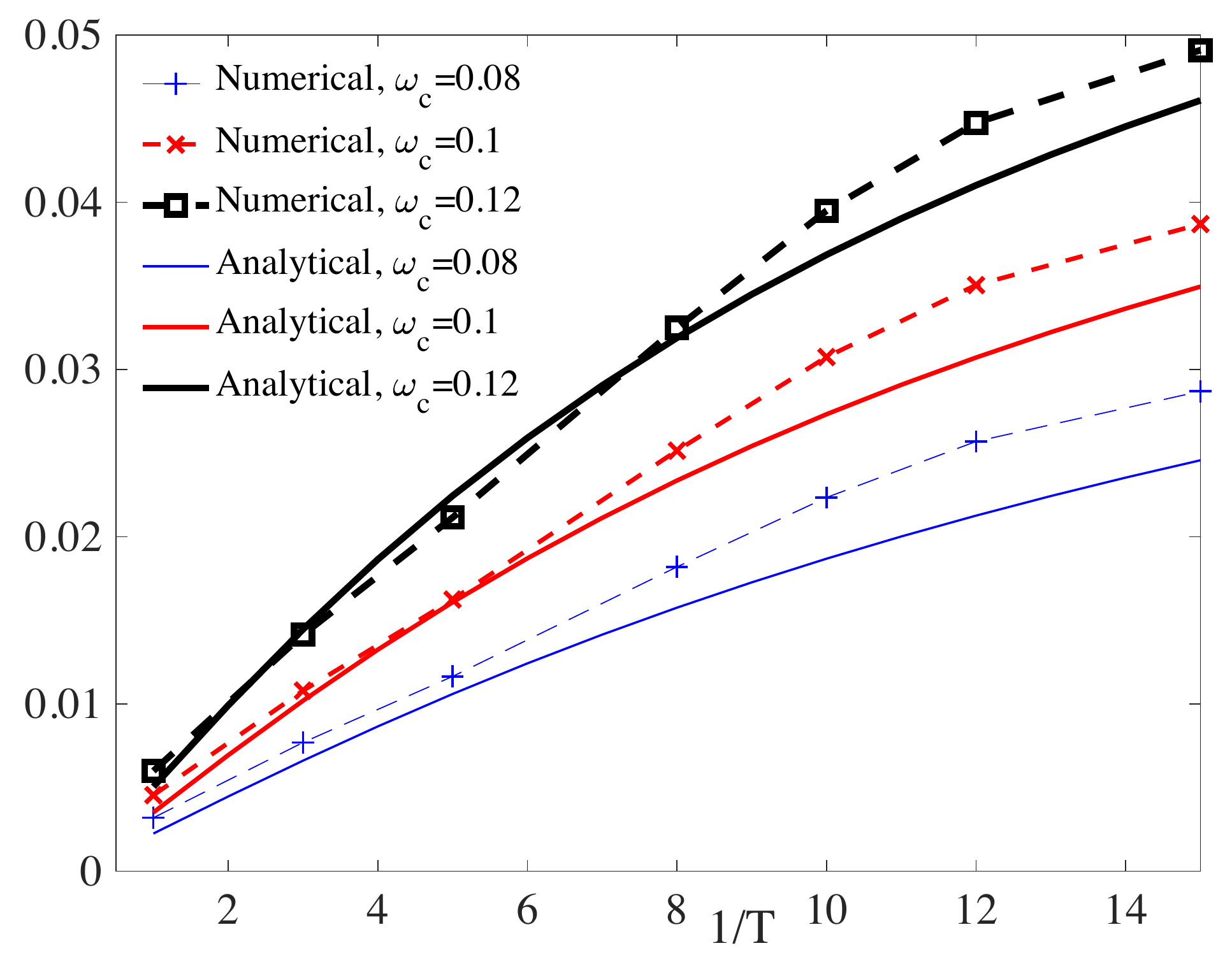}
\caption{The quantity $F(\omega_c, T) = \int_0^{\infty} d\omega
e^{-\omega/\omega_c} \omega \sigma(\omega,T)$ extracted from exact
diagonalization of Eq.~\ref{H} on a $3 \times 4$ lattice, with $g
= 1$, $M = 2$, $N = 3$, and a fixed particle number $N_p = 4$. The
solid lines are the plot of the same quantity calculated based on
the scaling function Eq.~\ref{theocond}. In the definition of
electric current we have also taken $N=3$, $M=2$, namely both the
nearest neighbor and second neighbor hopping will contribute to
conductivity. On this small system our data with a uniform
$\eta_{r,r'} = +1$ compares better with the analytical solution in
the large$-N,M$ limit.} \label{cond}
\end{center}
\end{figure}

We also studied the minimal and most realistic version of our
model, Eq.~\ref{Hs}, with exact diagonalization on a small $3
\times 4$ lattice with periodic boundary condition, and a fixed
particle number $N_p = 4$. With our numerical method, it is most
convenient to compare the quantity $F(\omega_c, T) =
\int_0^{\infty} d\omega e^{-\omega/\omega_c} \omega
\sigma(\omega,T)$ with the analytical result Eq.~\ref{theocond}.
We found that the case with a uniform choice $\eta_{r,r'} = +1$
compares better with the solution in the large$-N,M$ limit. The
general shape of the function $F(\omega_c, T)$ obtained
numerically is similar to the analytical expression in the
large$-N,M$ limit (Fig.~\ref{cond}), but further numerical
evidences are demanded for larger system sizes, for both choices
of $\eta_{r,r'}$.

The value of the DC conductivity is tunable by the parameter $t$
in the definition of the electric current (which is determined by
the size of the hopping term), and the overall energy scale $g$.
Thus the resistivity in the minimal version of our model can
easily exceed the Mott-Ioffe-Regel limit, $i.e.$ it can naturally
become the so-called ``bad metal", which is another puzzling
phenomenon observed in cuprate materials and has attracted a lot
of attentions~\cite{limit1,limit2,limit3}.

\subsection{Pairing instability and ``pseudogap"}

Besides hopping, we can also turn on other perturbations on
Eq.~\ref{Hs}. For example, we can turn on the following
perturbation on every link of the lattice: \beqn H_u = \sum_{< i,
j>} - \frac{u}{2M} \left( \Delta^\dagger_{i, j} \Delta_{i, j} +
\Delta_{i, j} \Delta^\dagger_{i, j} \right). \label{H1}\eeqn Here
$\Delta_{i, j} = \mathcal{J}_{\alpha\beta} c_{i,\alpha} c_{j,
\beta}$ is a Sp($M$) singlet pairing operator on a nearest
neighbor link $<i, j>$. This term can be reorganized into a
nearest neighbor density-density interaction and a Heisenberg
interaction using the Fierz identity of the symplectic Lie
algebra~\cite{readsachdev1991}.

This interaction term is marginal at the large$-N, M$ limit by
power-counting, again based on the fact that the fermion operator
has scaling dimension $1/4$, and in the large$-N,M$ limit all the
renormalization from Eq.~\ref{H} to this term is contained in the
renormalization of the fermion operator. In this limit, the RG
equation of $u$ can be computed through the standard loop diagram
in the same way as Ref.~\onlinecite{Xu2017}, using the fermion
Green's function in Eq.~\ref{green}: \beqn \frac{du}{d\ln l} =
\frac{u^{2}}{\sqrt{g^{2}\pi\cos\left(2\theta\right)}}. \label{rg}
\eeqn Thus the $u$ term is marginally relevant in this limit, and
it will likely lead to the fermion pairing instability just like
the BCS instability of the ordinary Fermi liquid.

$H_u$ and single particle hopping will compete with each other
under RG. $H_u$ will become nonperturbative at scales $T^\ast$:
\beqn T^\ast \sim g \exp \left(- \sqrt{\pi
\cos(2\theta)}\frac{g}{u} \right). \label{E} \eeqn Assuming the
single particle hopping becomes nonperturbative at scale $E_0$ (by
naive power-counting a single particle hopping is indeed relevant,
and will become nonperturbative at scale $E_0 \sim t^2/g$), Then
obviously there are two possible scenarios: If $E_0
> T^\ast$, the hopping term will dominate the low energy
physics and generate a Fermi sea. And at low energy the RG flow of
$u$ will be controlled by the standard RG equation of interactions
on the Fermi sea, and again $u$ will be marginally relevant and
lead to a pairing instability~\cite{shankar}. $H_u$ and the band
structure together will likely favor a $d-$wave
superconductor~\cite{dwave1,dwave2,dwave3} on the square lattice
near half-filling.

The possibility of $T^\ast > E_0$, $i.e.$ $u$ becomes
nonperturbative first under renomralization while lowering energy,
is even more interesting. Without single electron hopping, based
on the RG equation Eq.~\ref{rg} alone, one cannot determine the
pairing symmetry. In fact, in this case, while lowering
temperature (energy scale), before forming a superconductor with
global phase coherence, the system would favor to form Sp($M$)
spin singlet fermion pairings on as many nearest neighbor links as
possible. At half-filling, a generalization of the Rokhsar's
theorem~\cite{rokhsar} can be straightforwardly applied to our
case, and the ground states of Eq.~\ref{H1} in the large$-M$ limit
are all the ``dimerized" configurations with one quarter of the
links occupied by $M/2$ pairs of fermions that each forms a
Sp($M$) singlet~\footnote{Rokhsar's original theorem was proven
for spin systems instead of fermion systems. But this theorem was
formulated in the slave-fermion language, and the gauge constraint
on the slave-fermions becomes less and less important with
increasing $N$. In the large$-N$ limit, energetically the slave
fermions become physical fermions, because the gauge field
dynamics is completely suppressed in this limit.}. All these
dimerized configurations are degenerate in the large$-M$
limit~\cite{rokhsar}. Weak disorder and $1/M$ correction could
energetically select certain pattern of dimerization from the
extensively degenerate configurations, as was observed
experimentally~\cite{cupratedimer}. This state has a single
particle excitation gap which necessarily breaks a Sp($M$) singlet
on one of the links, but there is no global fermion-pair phase
coherence. This case could be identified as the pseudogap phase in
the cuprates phase diagram above the superconducting dome.

\begin{figure}[tbp]
\begin{center}
\includegraphics[width=240pt]{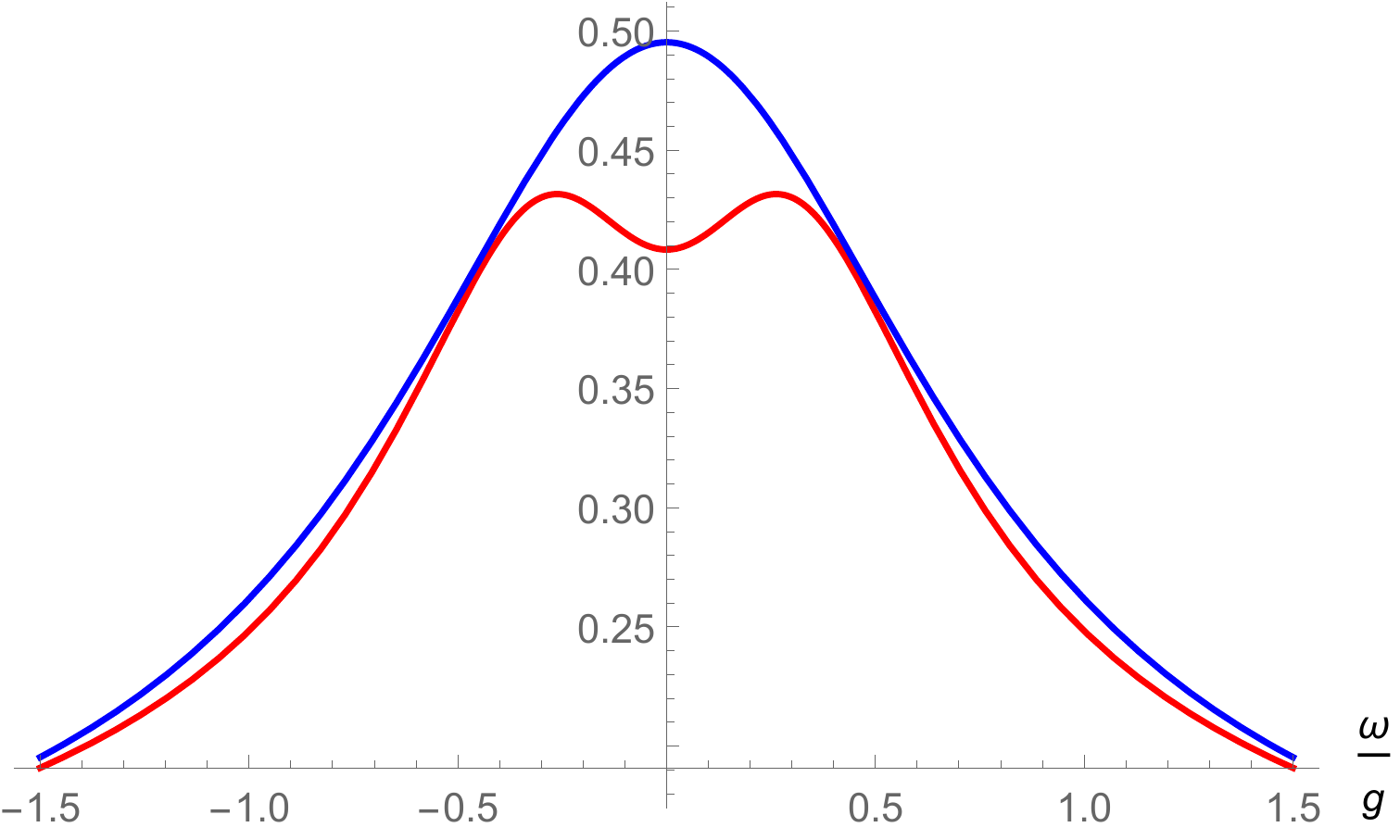}
\caption{The local density of states at half filling ($\theta =
0$) with $T > T^\ast$ and $\langle \Delta_{ij}\rangle = 0$ (blue
upper curve), and $T < T^\ast$ with nonzero $\langle
\Delta_{ij}\rangle$ (red lower curve). In the former case we have
chosen $g\beta = 2$; in the latter case we have chosen $g\beta =
4.5$ and $(u \Delta ) /(g M) = 0.15$ for illustration. }
\label{pseudo}
\end{center}
\end{figure}

The ``pseudogap" crossover temperature $T^\ast$ is given by
Eq.~\ref{E}, below which the system develops a nonzero expectation
value of $\langle \Delta_{ij} \rangle = \Delta$ on a maximal
possible number of links, based on our physical picture given
above. With a nonzero $\Delta$, for each pair of sites $i$ and $j$
coupled by the Sp$(M)$ singlet pair, we consider the perturbation
$\frac{u}{M} \Delta^\ast
\left(\mathcal{J}_{\gamma\delta}c_{i,\gamma}c_{j,\delta}\right)+H.c.$
to the original model Eq.~\ref{Ht2}. Let us consider two sites ($j
= 1,2$) connected by a dimer. We introduce a $2M$-component
fermion basis
$\Psi=\left(c_{1,\alpha}~,~c_{2,\alpha}^{\dagger}\right)^T$ and
the $2M\times 2M$ Green's function matrix $\mathcal{G}(\tau)
\equiv -\langle\mathbb{T}_\tau \Psi(\tau) \Psi(0)^\dag \rangle$.
To the first order of $\Delta$, the Green's function in the
imaginary-frequency domain is given by \beqn
\mathcal{G}^{-1}\left(i\omega_{n}\right)=\left[\begin{array}{cc}
G^{-1}\left(i\omega_{n}\right) & \frac{u}{M}\Delta\mathcal{J}\\
\frac{u}{M} \Delta^\ast \mathcal{J}^{\textrm{T}} &
-G^{-1}\left(-i\omega_{n}\right)
\end{array}\right],
\label{Nambu G} \eeqn where $G\left(i\omega_{n}\right)$ is the
original single fermion Green's function given by
Eq.~\ref{green},Eq.~\ref{green3}. By inverting Eq.~\ref{Nambu G},
we obtain the final Green's function $-\langle\mathbb{T}_\tau
c_{1,\alpha}( \tau ) c_{1,\beta}^\dag ( 0)\rangle$: \beqn
\frac{\delta_{\alpha\beta}}{G^{-1} \left(i\omega_{n}\right)+
\frac{u^{2}}{M^{2}}\left|\Delta\right|^{2}
G\left(-i\omega_{n}\right)}. \eeqn We can analytically continue
this expression to real frequency to obtain the retarded Green's
function on each site, whose imaginary part can be identified as
the local density of states (see Fig.~\ref{pseudo}), where a
``pseudogap" is manifest. In this calculation the Green's function
only depends on the amplitude of $ \langle \Delta_{ij} \rangle$,
thus even if the phase angle of $\langle \Delta_{ij} \rangle$ is
disordered the pseudogap in the local density of states is still
expected to exist.

A schematic global phase diagram with the parent strange metal
phase dominated by $H_s$, and the competition between
perturbations $H_u$ and single particle hopping parametrized by
$t$ is depicted in Fig.~\ref{pd}.

We must stress that all the analysis discussed in this section is
based on the physics of the tetrahedron model in the soluble
limit, which is identical to the disorder-averaged physics of the
SYK model. No matter how exactly the SYK physics is realized in
the real system, these analysis always applies. Our Eq.~\ref{Hs}
and Eq.~\ref{H} only give one possible realization of these
physics. Very similar physics can be realized in another model
discussed in the following section.

\section{Another possible model}

Another model which is slightly less natural but probably leads to
very similar physics is also worth discussion. Again, the most
important term (but not the only term) of the Hamiltonian reads
\beqn && H = \sum_j H_j, \cr\cr && H_j = U \hat{n}_{j}^2 +
\sum_{\hat{e} = \hat{e}_1, \hat{e}_2}J \left( \vec{S}_{j} \cdot
\vec{S}_{j + \hat{e}} - \frac{1}{4} \hat{n}_{j}\hat{n}_{j +
\hat{e}} \right) \cr\cr &-& K \left( \epsilon_{\alpha\beta}
\epsilon_{\gamma\sigma} c^\dagger_{j, \alpha} c^\dagger_{j +
\hat{e}_1 + \hat{e}_2, \beta} c_{j+\hat{e}_2, \gamma} c_{j +
\hat{e}_1, \sigma} + H.c. \right), \label{Hs2} \eeqn where
$\hat{e}_1 = \hat{x} + \hat{y}$, and $\hat{e}_2 = \hat{x} -
\hat{y}$. This term has no interaction between sublattice A and B
yet, and like before we will consider the single particle hoppings
and interactions that mix the two sublattices as perturbations.

The advantage of this model is that, we no longer needs a
large$-N$ generalization of the hopping term. The ordinary nearest
neighbor hopping bridges the two sublattices, $i.e.$ it bridges
two ``SYK-clusters", similar to the previously studied coupled SYK
cluster models~\cite{altman,zhai2017}. The nearest neighbor
hopping with coefficient $t$ is a relevant perturbation based on
the scaling dimension of the fermion operator $\Delta[c_j] = 1/4$
in the soluble limit. The scaling dimension of $t$ is $\Delta[t] =
1/2$. Thus with the perturbation of the nearest neighbor hopping,
we expect the large$-N,M$ solution of the tetrahedron model to be
applicable roughly to the energy window $(t^2/g, \ g)$, and within
this window the longitudinal conductivity $\sigma(\omega, T)$
takes the same form as the previous case. Other analysis like the
perturbation of $H_u$ (Eq.~\ref{H1}) and pairing instability
remains unchanged compared with the last model we considered.

\section{Summary and Discussion}

In this work we proposed two strongly interacting electron models
on the square lattice, with one orbital per unit cell. And we
demonstrated that in certain limit these models mimic the behavior
of the ``tetrahedron" tensor model, and hence can be solved. The
physics in this limit is consistent with the main phenomenology of
the strange metal non fermi liquid phase observed in the cuprates.
We argue that away from this exactly soluble limit, there is still
a finite energy window where the solution is applicable. We then
checked our predictions numerically by exactly diagonalizing the
minimal version of the proposed Hamiltonian (which is away from
the soluble limit and hence takes a realistic form) on a small
lattice. We also discussed effects of perturbations including the
single particle hopping, and argued that depending on the
competition between two perturbations, the system can develop
either a $d-$wave superconductor, or a ``pseudogap" phase at low
temperature.

More numerical effort is demanded in the future to further analyze
both our models Eq.~\ref{Hs}, Eq.~\ref{Hs2}. Also, more
predictions on thermodynamics and transport can be made below the
crossover temperature $T^\ast$ where the system enters the
pseudogap phase driven by $H_u$. The exact phase boundaries in the
phase diagram Fig.~\ref{pd} also needs further detailed
calculations. In this work we have treated single particle hopping
as a perturbation on top of the SYK-like physics. A complete
treatment of the interaction term Eq.~\ref{Hs}, Eq.~\ref{Hs2}
together with a single particle hopping is demanded in the future
in order to study the momentum space structure of our theory. We
will leave these open questions to future studies.

Chen is supported by a postdoctoral fellowship from the Gordon and
Betty Moore Foundation, under the EPiQS initiative, Grant
GBMF4304, at the Kavli Institute for Theoretical Physics. Xu is
supported by the David and Lucile Packard Foundation and NSF Grant
No. DMR-1151208. We acknowledge support from the Center for
Scientific Computing from the CNSI, MRL: an NSF MRSEC
(DMR1121053). We also thank Leon Balents, Matthew P. A. Fisher,
Steven A. Kivelson, Subir Sachdev and T. Senthil for very helpful
discussions.

\bibliography{SYK}

\end{document}